\documentclass[10pt,conference,a4paper]{IEEEtran}
\usepackage{amsmath}
\usepackage{amssymb}
\usepackage{amsfonts}
\usepackage{amscd}
\usepackage{mathrsfs}
\usepackage[final]{graphicx}
\usepackage{color}
\usepackage{url}
\usepackage{verbatim}

\newtheorem{eg}{Example}

\begin{document}

\title{OFDM based Distributed Space Time Coding for Asynchronous Relay Networks}

\author{
\authorblockN{G. Susinder Rajan}
\authorblockA{ECE Department \\
Indian Institute of Science\\
Bangalore 560012, India\\
susinder@ece.iisc.ernet.in}
\and
\authorblockN{B. Sundar Rajan}
\authorblockA{ECE Department\\
Indian Institute of Science\\
Bangalore 560012, India\\
bsrajan@ece.iisc.ernet.in}
}

\maketitle

\begin{abstract}
Recently Li and Xia have proposed a transmission scheme for wireless relay networks based on the Alamouti space time code and orthogonal frequency division multiplexing to combat the effect of timing errors at the relay nodes. This transmission scheme is amazingly simple and achieves a diversity order of two for any number of relays. Motivated by its simplicity, this scheme is extended to a more general transmission scheme that can achieve full cooperative diversity for any number of relays. The conditions on the distributed space time block code (DSTBC) structure that admit its application in the proposed transmission scheme are identified and it is pointed out that the recently proposed full diversity four group decodable DSTBCs from precoded co-ordinate interleaved orthogonal designs and extended Clifford algebras satisfy these conditions. It is then shown how differential encoding at the source can be combined with the proposed transmission scheme to arrive at a new transmission scheme that can achieve full cooperative diversity in asynchronous wireless relay networks with no channel information and also no timing error knowledge at the destination node. Finally, four group decodable distributed differential space time block codes applicable in this new transmission scheme for power of two number of relays are also provided.
\end{abstract}

\section{Introduction}
\label{sec1}

Coding for cooperative wireless relay networks has attracted considerable attention recently. Distributed space time coding was proposed as a coding strategy to achieve full cooperative diversity in \cite{JiH} assuming that the signals from all the relay nodes arrive at the destination at the same time. But this assumption is not close to practicality since the relay nodes are geographically distributed. In \cite{GuX}, a transmission scheme based on orthogonal frequency division multiplexing (OFDM) at the relay nodes was proposed to combat the timing errors at the relays and a high rate space time block code (STBC) construction was also provided. However, the maximum likelihood (ML) decoding complexity for this scheme is prohibitively high especially for the case of large number of relays. Several other works in the literature propose methods to combat the timing offsets but most of them are based on decode and forward at the relay node and moreover fail to address the decoding complexity issue. In \cite{LiX}, a simple transmission scheme to combat timing errors at the relay nodes was proposed. This scheme is particularly interesting because of its associated low ML decoding complexity. In this scheme, OFDM is implemented at the source node and time reversal/conjugation is performed at the relay nodes on the received OFDM symbols. The received signals at the destination after OFDM demodulation are shown to have the Alamouti code structure and hence single symbol maximum likelihood (ML) decoding can be performed. However, the Alamouti code is applicable only for the case of two relay nodes and for larger number of relays, the authors of \cite{LiX} propose to cluster the relay nodes and employ Alamouti code in each cluster. But this clustering technique provides diversity order of only two and fails to exploit the full cooperative diversity equal to the number of relay nodes.

The main contributions of this paper are as follows.
\begin{itemize}
\item The Li-Xia transmission scheme is extended to a more general transmission scheme that can achieve full asynchronous cooperative diversity for any number of relays.
\item The conditions on the distributed STBC (DSTBC) structure that admit its application in the proposed transmission scheme are identified. The recently proposed full diversity four group decodable DSTBCs in \cite{RaR1} for synchronous wireless relay networks are found to satisfy the required conditions for application in the proposed transmission scheme.
\item It is shown how differential encoding at the source node can be combined with the proposed transmission scheme to arrive at a transmission scheme that can achieve full asynchronous cooperative diversity in the absence of channel knowledge and in the absence of knowledge of the timing errors of the relay nodes. Moreover, an existing class of four group decodable distributed differential STBCs \cite{RaR2} for synchronous relay networks with power of two number of relays is shown to be applicable in this setting as well.
\end{itemize}

\subsection{Organization of the paper}
\label{subsec1_1}
In Section \ref{sec2}, the basic assumptions on the relay network model are given and the proposed transmission scheme is described. Section \ref{sec2} also provides four group decodable DSTBCs that achieve full asynchronous cooperative diversity in the proposed transmission scheme for arbitrary number of relays. Section \ref{sec3} briefly explains how differential encoding at the source node can be combined with the proposed transmission scheme and four group decodable distributed differential STBCs applicable in this scenario are also proposed. Simulation results and discussion on further work comprise Sections \ref{sec4} and \ref{sec5} respectively.\\
~\\
\noindent
\textbf{Notation:}
$\mathbf{I_m}$ denotes an $m\times m$ identity matrix and $\mathbf{0}$ denotes an all zero matrix of appropriate size. For a set $A$, the cardinality of $A$ is denoted by $|A|$. A null set is denoted by $\phi$. For a matrix, $(.)^T$, $(.)^*$ and $(.)^H$ denote transposition, conjugation and conjugate transpose operations respectively. For a complex number, $(.)_I$ and $(.)_Q$ denote its in-phase and quadrature-phase parts respectively.

\section{Relay network model assumptions and the proposed transmission scheme}
\label{sec2}

In this section, the basic relay network model assumptions are given and the proposed transmission scheme is described. The proposed transmission scheme can achieve full asynchronous cooperative diversity for arbitrary number of relays and is an extension of the Li-Xia transmission scheme\cite{LiX}. This nontrivial extension is based on analyzing the sufficient conditions required on the structure of STBCs which admit application in the Li-Xia transmission scheme.

\begin{figure}[h]
\centering
\input{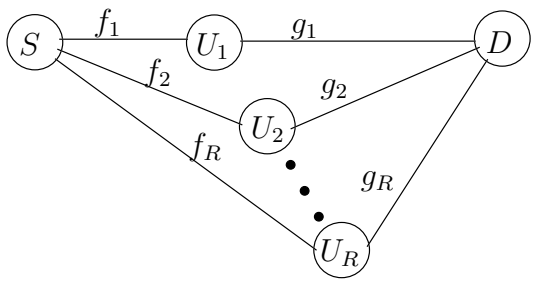}
\caption{Asynchronous wireless relay network}
\label{fig_network}
\end{figure}

\subsection{Network model assumptions}
\label{subsec2_1}
Consider a network with one source node, one destination node and $R$ relay nodes $U_1, U_2, \dots, U_R$. This is depicted in Fig. \ref{fig_network}. Every node is assumed to have only a single antenna and is half duplex constrained.  The channel gain between the source and the $i$-th relay $f_i$ and that between the $j$-th relay and the destination $g_j$ are assumed to be quasi-static, flat fading and modeled by independent and complex Gaussian distributed random variables with mean zero and unit variance. The transmission of information from the source node to the destination node takes place in two phases. In the first phase, the source broadcasts the information to the relay nodes using OFDM. The relay nodes receive the faded and noise corrupted OFDM symbols, process them and transmit them to the destination. The relay nodes are assumed to have perfect carrier synchronization. The overall relative timing error of the signals arrived at the destination node from the $i$-th relay node is denoted by $\tau_i$. Without loss of generality, it is assumed that $\tau_1=0$, $\tau_{i+1}\geq \tau_i,i=1,\dots,R-1$. The destination node is assumed to have the knowledge of all the channel fading gains $f_i,g_j, i,j=1,\dots,R$ and the relative timing errors $\tau_i,i=1,\dots,R$.

\subsection{Transmission by the source node}
\label{subsec2_2}

The source takes $RN$ complex symbols $x_{i,j, 0\leq i\leq N-1, j=1,2,\dots,R}$ and forms $R$ blocks of data denoted by $\mathbf{x_j}=\left[\begin{array}{cccc}x_{0,j} & x_{1,j} & \dots & x_{N-1,j}\end{array}\right]^T, j=1,2,\dots,R$. Of these $R$ blocks, $M$ of them are modulated by $N$-point IDFT and the remaining $R-M$ blocks are modulated by $N$-point DFT. Without loss of generality, let us assume that the first $M$ blocks are modulated by $N$-point IDFT. Then a CP of length $l_{cp}$ is added to each block, where $l_{cp}$ is chosen to be not less than the maximum of the overall relative timing errors of the signals arrived at the destination node from all the relay nodes. The resulting $R$ OFDM symbols denoted by $\mathbf{\bar{x}_1},\mathbf{\bar{x}_2},\dots,\mathbf{\bar{x}_R}$ each consisting of $L_s=N+l_{cp}$ complex numbers are broadcasted to the $R$ relays using a fraction $\pi_1$ of the total average $P$.

\subsection{Processing at the relay nodes}
\label{subsec2_3}

If the channel fade gains are assumed to be constant for $2R$ OFDM symbol intervals, the received signals at the $i$-th relay during the $j$-th OFDM symbol duration is given by

$$
\mathbf{r_{i,j}}=\sqrt{\pi_1P}f_i\mathbf{\bar{x}_j}+\mathbf{\bar{v}_{i,j}}
$$

\noindent where, $\mathbf{\bar{v}_{i,j}}$ is the AWGN at the $i$-th relay node during the $j$-th OFDM symbol duration. The relay nodes process and transmit the received noisy signals as shown in Table \ref{table_proposed} using a fraction $\pi_2$ of total power $P$ (appropriate scaling of received OFDM symbols is assumed). Note from Table \ref{table_proposed} that time reversal is done during the last $R-M$ OFDM symbol durations. We would like to emphasize that in general time reversal could be implemented in any $R-M$ of the total $R$ OFDM symbol durations. Now, $\mathbf{t_{i,j}}\in\left\{\mathbf{0},\pm \mathbf{r_{i,j}}, j=1,\dots,R\right\}$ with the constraint that the $i$-th relay should not be allowed to transmit any element from the the following set:

{\small
\begin{equation}
\label{eqn_constraints}
\begin{array}{l}
\left\{\pm\mathbf{r_{i,j}}^*, j=1,\dots,M\right\}
\cup\left\{\pm\zeta(\mathbf{r_{i,j}}),j=1,\dots,M\right\}\\
\cup\left\{\pm\mathbf{r_{i,j}},j=M+1,\dots,R\right\}
\cup\left\{\pm\zeta(\mathbf{r_{i,j}}^*), j=M+1,\dots,R\right\}.
\end{array}
\end{equation}
}

\begin{table*}
\caption{Proposed transmission scheme}
\label{table_proposed}
{\footnotesize
\begin{center}
\begin{tabular}{|c|c|c|c|c|c|c|}
\hline
OFDM Symbol & $U_1$ & $\dots$ & $U_{M}$ & $U_{M+1}$ & $\dots$ & $U_R$\\
\hline
$1$ & $\mathbf{t_{1,1}}$ & $\dots$ & $\mathbf{t_{M,1}}$ & $\mathbf{t_{M+1,1}}^*$ & $\dots$ & $\mathbf{t_{R,1}}^*$\\
\hline
$\vdots$ & $\vdots$ & $\vdots$ & $\vdots$ & $\vdots$ & $\vdots$ & $\vdots$\\
\hline
$M$ & $\mathbf{t_{1,M}}$ & $\dots$ & $\mathbf{t_{M,M}}$ & $\mathbf{t_{M+1,M}}^*$ & $\dots$ & $\mathbf{t_{R,M}}^*$\\
\hline
$M+1$ & $\zeta(\mathbf{t_{1,M+1}})$ & $\dots$ & $\zeta(\mathbf{t_{M,M+1}})$ & $\zeta(\mathbf{t_{M+1,M+1}}^*)$ & $\dots$ & $\zeta(\mathbf{t_{R,M+1}}^*)$\\
\hline
$\vdots$ & $\vdots$ & $\vdots$ & $\vdots$ & $\vdots$ & $\vdots$ & $\vdots$\\
\hline
$R$ & $\zeta(\mathbf{t_{1,R}})$ & $\dots$ & $\zeta(\mathbf{t_{M,R}})$ & $\zeta(\mathbf{t_{M,R}}^*)$ & $\dots$ & $\zeta(\mathbf{t_{R,R}}^*)$\\
\hline
\end{tabular}
\end{center}
}
\end{table*}

\subsection{Decoding at the destination}
\label{subsec2_4}

The destination removes the CP for the first $M$ OFDM symbols and implements the following for the remaining OFDM symbols:

\begin{enumerate}
\item Remove the CP to get a $N$-point vector
\item Shift the last $l_{cp}$ samples of the $N$-point vector as the first $l_{cp}$ samples.
\end{enumerate}

DFT is then applied on the resulting $R$ vectors. Let the received signals for $R$ consecutive OFDM blocks after CP removal and DFT transformation be denoted by $\mathbf{y_j}=\left[\begin{array}{cccc}y_{0,j} & y_{1,j} & \dots & y_{N-1,j}\end{array}\right]^T, j=1,2,\dots,R$. Let $\mathbf{w_i}=(w_{k,i}), i=1,\dots,R$ represent the AWGN at the destination node and let $\mathbf{v_{i,j}}$ denote the DFT of $\mathbf{\bar{v}_{i,j}}$. Let $\mathbf{s_k}=\left[\begin{array}{cccc}x_{k,1} & x_{k,2} & \dots & x_{k,R}\end{array}\right]^T, k=0,1,\dots,N-1$. Now using the following identities,

\begin{equation}
\label{eqn_identities}
\begin{array}{lcl}
(\mathrm{DFT}(\mathbf{x}))^*&=&\mathrm{IDFT}(\mathbf{x}^*)\\ (\mathrm{IDFT}(\mathbf{x}))^*&=&\mathrm{DFT}(\mathbf{x}^*)\\  \mathrm{DFT}(\zeta(\mathrm{DFT}(\mathbf{x})))&=&\mathbf{x}
\end{array}
\end{equation}

\noindent we get in each sub carrier $k,0\leq k\leq N-1$:

\begin{equation}
\label{eqn_sys_model}
\mathbf{y_{k}}=\left[\begin{array}{cccc}y_{k,1} & y_{k,2} & \dots & y_{k,R}\end{array}\right]^T=\sqrt{\frac{\pi_1\pi_2P^2}{\pi_1P+1}}\mathbf{X_kh_k}+\mathbf{n_k}
\end{equation}

\noindent where,
\begin{equation}
\label{eqn_conj_linear}
\mathbf{X_k}=\left[\begin{array}{cccccc}\mathbf{A_1s_k} & \dots & \mathbf{A_{M}s_k} & \mathbf{A_{M+1}}\mathbf{s_k}^* & \dots \mathbf{A_{R}}\mathbf{s_k}^* \end{array}\right]
\end{equation}

\noindent for some square real matrices $\mathbf{A_i}, i=1,\dots,R$ having the property that any row of $\mathbf{A_i}$ has only one nonzero entry. If $u_k^{\tau_i}=e^{-\frac{i2\pi k\tau_i}{N}}$, then
\begin{equation}
\label{eqn_channel}
\mathbf{h_k}=\left[\begin{array}{c} f_1g_1\\ u_k^{\tau_2}f_2g_2\\ \dots\\
u_k^{\tau_{M}}f_{M}g_{M}\\
u_k^{\tau_{M+1}}f_{M+1}^*g_{M+1}\\ \dots\\
u_k^{\tau_R}f_R^*g_R\end{array}\right]
\end{equation}
\noindent is the equivalent channel matrix for the $k$-th sub carrier. The equivalent noise vector is given by

$$
\begin{array}{rl}
\mathbf{n_k}=&\sqrt{\frac{\pi_2P}{\pi_1P+1}}\left[\begin{array}{c}\delta_1\sum_{i=1}^{R}sgn(\mathbf{t_{i,1}})\mathbf{\hat{v}_{i,1}}(k)g_iu_k^{\tau_i}\\
\delta_2\sum_{i=1}^{R}sgn(\mathbf{t_{i,2}})\mathbf{\hat{v}_{i,2}}(k)g_iu_k^{\tau_i}\\
\vdots\\
\delta_R\sum_{i=1}^{R}sgn(\mathbf{t_{i,R}})\mathbf{\hat{v}_{i,R}}(k)g_iu_k^{\tau_i}
\end{array}\right]\\
&+\left[\begin{array}{c}w_{k,1}\\ w_{k,2}\\ \dots\\  w_{k,R}\end{array}\right]
\end{array}
$$

\noindent where, $sgn(\mathbf{t_{i,j}})=\left\{\begin{array}{l} 1~~\mathrm{if}~ \mathbf{t_{i,j}}\in\left\{\mathbf{r_{i,j}}, j=1,\dots,R\right\}\\
-1~~\mathrm{if}~\mathbf{t_{i,j}}\in\left\{-\mathbf{r_{i,j}}, j=1,\dots,R\right\}\\
0~~\mathrm{if}~\mathbf{t_{i,j}}=\mathbf{0}\end{array}\right.$ and $\mathbf{\hat{v}_{i,m}}=\left\{\begin{array}{l} \pm\mathbf{v_{i,j}}~\mathrm{if}~i\leq M~\mathrm{and}~\mathbf{t_{i,m}}=\pm\mathbf{r_{i,j}}\\
\pm\mathbf{v_{i,j}}^*~\mathrm{if}~i>M~\mathrm{and}~\mathbf{t_{i,m}}=\pm\mathbf{r_{i,j}}\end{array}\right.$. The $\delta_i$'s are simply scaling factors to account for the correct noise variance due to possible zeros in the relay transmissions.

ML decoding of $\mathbf{X_k}$ can be done from \eqref{eqn_sys_model} by choosing that codeword which minimizes $\parallel\Omega^{-\frac{1}{2}}(\mathbf{y_k}-\mathbf{X_k}\mathbf{h_k})\parallel_F^2$, where $\Omega$ is the covariance matrix of $\mathbf{n_k}$ and $\parallel .\parallel_F$ denotes the Frobenius norm. Essentially, the proposed transmission scheme implements a space time code having a special structure in each sub carrier. Now if the DSTBC $\mathbf{X_k}$ satisfies the rank criteria (difference of any two codeword matrices has full rank), then it can be proved on similar lines as in \cite{JiH} that full asynchronous cooperative diversity equal to $R$ is achieved.

\subsection{Full diversity four group decodable distributed space time codes}
\label{subsec2_5}

In this subsection, we analyze the structure of the space time code required for implementing in the proposed transmission scheme. Note from \eqref{eqn_conj_linear} that the DSTBC should have the property that any column should have only the complex symbols or only their conjugates. We refer to this property as \textit{conjugate linearity} property\cite{RaR1}. But conjugate linearity alone is not enough for a STBC to qualify for implementation in the proposed transmission scheme. Note from Table \ref{table_proposed} that time reversal is implemented for certain OFDM symbol durations by all the relay nodes. Observe that this put together with the constraints in \eqref{eqn_constraints} demands a certain row structure on the STBC. We now provide a set of sufficient conditions that are required on the row structure of conjugate linear STBCs. First let us partition the complex symbols appearing in the $i$-th row into two sets- one set $P_i$ containing those complex symbols which appear without conjugation and another set $P_i^c$ which contains those complex symbols which appear with conjugation in the $i$-th row. If the following sufficient conditions are satisfied by a conjugate linear STBC, then it can be shown that there exists an assignment of time reversal OFDM symbol durations together with an appropriate choice of $M$ and relay node processing such that the desired conjugate linear STBC form is obtained in every sub carrier at the destination node.

\begin{equation}
\label{eqn_row_property}
\begin{array}{c}
P_i\cap P_i^c=\phi,~\forall~i=1,\dots,R\\
|P_i|=|P_i^c|,~ \forall~ i=1,\dots,R\\
P_i\cap P_j\in\left\{\phi, P_i, P_j\right\},~\forall~ i\neq j.
\end{array}
\end{equation}

Now that for the case of the Alamouti code\cite{LiX}, $P_1=P_2^c=\left\{x_{k,1}\right\}$, $P_2=P_1^c=\left\{x_{k,2}\right\}$ and hence it satisfies the conditions in \eqref{eqn_row_property}. Recently three new classes of full diversity, four group decodable DSTBCs for any number of relays were reported in \cite{RaR1} for synchronous relay networks. These codes are conjugate linear and moreover since they are four group decodable, the associated real symbols in these STBCs can be partitioned equally into four groups and the ML decoding can be done for the real symbols in a group independently of the real symbols in the other groups. Thus the ML decoding complexity of these codes is significantly less compared to all other distributed space time codes known in the literature. In this paper, we show that the codes reported in \cite{RaR1} satisfy the conditions in \eqref{eqn_row_property} and are thus suitable to be applied in the proposed transmission scheme. Due to space limitations this is illustrated using the following example.

\begin{eg}
\label{eg_5relay}
Let us take $R=5$, for which the DSTBC in \cite{RaR1} is obtained by taking a DSTBC for $6$ relays and dropping one column. It is given by
$$
\left[\begin{array}{ccccc}
x_{k,1} & -x_{k,2}^* & 0 & 0 & 0\\
x_{k,2} & x_{k,1}^* & 0 & 0 & 0\\
0 & 0 & x_{k,3} & -x_{k,4}^* & 0\\
0 & 0 & x_{k,4} & x_{k,3}^* & 0\\
0 & 0 & 0 & 0 & x_{k,5}\\
0 & 0 & 0 & 0 & x_{k,6}
\end{array}\right]
$$
\noindent for which $P_1=P_2^c=\left\{x_{k,1}\right\}$, $P_2=P_1^c=\left\{x_{k,2}\right\}$, $P_3=P_4^c=\left\{x_{k,3}\right\}$, $P_4=P_3^c=\left\{x_{k,4}\right\}$, $P_5=\left\{x_{k,5}\right\}$, $P_6=\left\{x_{k,6}\right\}$ and $P_5^c=P_6^c=\phi$. At the source, we choose $\mathbf{\bar{x}_1}=\mathrm{IDFT}(\mathbf{x_1})$, $\mathbf{\bar{x}_2}=\mathrm{DFT}(\mathbf{x_2})$,
$\mathbf{\bar{x}_3}=\mathrm{IDFT}(\mathbf{x_3})$,
$\mathbf{\bar{x}_4}=\mathrm{DFT}(\mathbf{x_4})$,
$\mathbf{\bar{x}_5}=\mathrm{IDFT}(\mathbf{x_5})$ and
$\mathbf{\bar{x}_6}=\mathrm{DFT}(\mathbf{x_6})$. The $5$ relays process the received OFDM symbols as shown in Table \ref{table_5relay}.

\begin{table}[h]
\caption{Transmission scheme for $5$ relays}
\label{table_5relay}
\begin{center}
{\footnotesize
\begin{tabular}{|c|c|c|c|c|c|}
\hline
OFDM & $U_1$ & $U_2$ & $U_3$ & $U_4$ & $U_5$\\
Symbol & & & & &\\
\hline
$1$ & $\mathbf{r_{1,1}}$ & $-\mathbf{r_{2,2}}^*$ & $\mathbf{0}$ & $-\mathbf{0}$ & $\mathbf{0}$\\
\hline
$2$ & $\zeta(\mathbf{r_{1,2}})$ & $\zeta(\mathbf{r_{2,1}}^*)$ &
$-\mathbf{0}$ & $-\mathbf{0}$ & $\mathbf{0}$\\
\hline
$3$ & $\mathbf{0}$ & $\mathbf{0}$ & $\mathbf{r_{3,3}}$ & $-\mathbf{r_{4,4}}^*$ & $\mathbf{0}$\\
\hline
$4$ & $\mathbf{0}$ & $\mathbf{0}$ &
$-\zeta(\mathbf{r_{3,2}})$ & $\zeta(\mathbf{r_{4,1}}^*)$ & $\mathbf{0}$\\
\hline
$5$ & $\mathbf{0}$ & $\mathbf{0}$ & $\mathbf{0}$ & $\mathbf{0}$ & $\mathbf{r_{5,5}}$\\
\hline
$6$ & $\mathbf{0}$ & $\mathbf{0}$ & $\mathbf{0}$ & $\mathbf{0}$ &
$-\zeta(\mathbf{r_{5,6}})$\\
\hline
\end{tabular}
}
\end{center}
\end{table}

This code is $3$ real symbol decodable and achieves full diversity for appropriately signal sets \cite{RaR1}.
\end{eg}

\section{Transmission Scheme for Noncoherent Asynchronous Relay Networks}
\label{sec3}

In this section, it is shown how differential encoding can be combined with the proposed transmission scheme described in Section \ref{sec3}. Then the codes in \cite{RaR2} are proposed for application in this setting.

For the proposed transmission scheme in Section \ref{sec3}, at the end of one transmission frame,  we have in the $k$-th sub carrier $\mathbf{y_k}=\sqrt{\frac{\pi_1\pi_2P^2}{\pi_1P+1}}\mathbf{X_k}\mathbf{h_k}+\mathbf{n_k}$. Note that the channel matrix $\mathbf{h_k}$ as shown in \eqref{eqn_channel} depends on $f_i,g_i,\tau_i, i=1,\dots,R$. Thus the destination node needs to have the knowledge of these values in order to perform ML decoding.

Now using differential encoding ideas which were proposed in \cite{KiR,OgH2,JiJ} for non-coherent communication in synchronous relay networks, we combine them with the proposed asynchronous transmission scheme. Supposing the channel remains approximately constant for two transmission frames, then differential encoding can be done at the source node in each sub carrier $0\leq k\leq N-1$ as follows:
$$
\mathbf{s_k^0}=\left[\begin{array}{cccc}\sqrt{R} & 0 & \dots & 0\end{array}\right]^T,~ \mathbf{s_k^t}=\frac{1}{a_t-1}\mathbf{C_t}\mathbf{s_k^{t-1}}, \mathbf{C_t}\in\mathscr{C}
$$
where, $s_k^{i}$ denotes the vector of complex symbols transmitted by the source during the $i$-th transmission frame in the $k$-th sub carrier and $\mathscr{C}$ is the codebook used by the source which consists of scaled unitary matrices \mbox{$\mathbf{C_t}^H\mathbf{C_t}=a_t^2\mathbf{I}$} such that $\mathrm{E}[a_t^2]=1$. If for all $\mathbf{C\in\mathscr{C}}$,

$$
\begin{array}{l}
\mathbf{C}\mathbf{A_i}=\mathbf{A_i}\mathbf{C}, i=1,\dots,M~\mathrm{and}\\ \mathbf{C}\mathbf{A_i}=\mathbf{A_i}\mathbf{C}^*, i=M+1,\dots,R
\end{array}
$$
then we have:

\begin{equation}
\mathbf{y_k^t}=\frac{1}{a_{t-1}}\mathbf{C_t}\mathbf{y_k^{t-1}}+(\mathbf{n_k^t}-\frac{1}{a_t-1}\mathbf{C_{t}}\mathbf{n_k^{t-1}})
\end{equation}
from which $\mathbf{C_t}$ can be decoded as \mbox{$\mathbf{\hat{C}_t}=\arg\min_{\mathbf{C_t\in\mathscr{C}}}\parallel\mathbf{y_k^t}-\frac{1}{a_{t-1}}\mathbf{C_t}\mathbf{y_k^{t-1}}\parallel_F^2$} in each sub carrier $0\leq k\leq N-1$.

Note that this decoder does not require the knowledge of $f_i,g_i,\tau_i, i=1,\dots R$ at the destination. However it is important to note that the knowledge of the maximum of the relative timing errors is needed to decide the length of CP.

It turns out that the four group decodable distributed differential space time codes constructed in \cite{RaR2} for synchronous relay networks with power of two number of relays meet all the requirements for use in the proposed transmission scheme as well. The following example illustrates this fact.

\begin{eg}
\label{eg_4relay_diff}
Let $R=4$. The codebook at the source is given by \\\mbox{$\mathscr{C}=\left\{\sqrt{\frac{1}{4}}\left[\begin{array}{cccc}
z_1 & z_2 & -z_3^* & -z_4^*\\
z_2 & z_1 & -z_4^* & -z_3^*\\
z_3 & z_4 & z_1^* & z_2^*\\
z_4 & z_3 & z_2^* & z_1^*
 \end{array}\right]\right\}$} where $\left\{z_{1I},z_{2I}\right\}$, $\left\{z_{1Q},z_{2Q}\right\}$, $\left\{z_{3I},z_{4I}\right\}$, $\left\{z_{3Q},z_{4Q}\right\}\in\mathbb{S}$ and \mbox{$\mathbb{S}=\left\{\left[\begin{array}{c}\frac{1}{\sqrt{3}}\\0\end{array}\right],\left[\begin{array}{c}-\frac{1}{\sqrt{3}}\\0\end{array}\right],\left[\begin{array}{c}0\\ \sqrt{\frac{5}{3}}\end{array}\right],\left[\begin{array}{c}0\\ -\sqrt{\frac{5}{3}}\end{array}\right] \right\}$}. Differential encoding is done at the source node for each sub carrier $0\leq k\leq N-1$ as follows:
$$
\mathbf{s_k^0}=\left[\begin{array}{cccc}\sqrt{R} & 0 & \dots & 0\end{array}\right]^T,~ \mathbf{s_k^t}=\frac{1}{a_t-1}\mathbf{C_t}\mathbf{s_k^{t-1}}, \mathbf{C_t}\in\mathscr{C}.
$$
Once we get $\mathbf{s_k^t},k=0,\dots,N-1$ from the above equation, the $N$ length vectors $\mathbf{x_i}, i=1,\dots,R$ can be obtained. Then IDFT/DFT is applied on these vectors and broadcasted to the relay nodes according to: $\mathbf{\bar{x}_1}=\mathrm{IDFT}(\mathbf{x_1})$, $\mathbf{\bar{x}_2}=\mathrm{IDFT}(\mathbf{x_2})$, $\mathbf{\bar{x}_3}=\mathrm{DFT}(\mathbf{x_3})$ and $\mathbf{\bar{x}_4}=\mathrm{DFT}(\mathbf{x_4})$. 

\begin{table}[h]
\caption{Transmission scheme for $4$ relays}
\label{table_4relay}
\begin{center}
\begin{tabular}{|c|c|c|c|c|}
\hline
OFDM & $U_1$ & $U_2$ & $U_3$ & $U_4$\\
Symbol & & & &\\
\hline
$1$ & $\mathbf{r_{1,1}}$ & $\mathbf{r_{2,2}}$ & $-\mathbf{r_{3,3}}^*$ & $-\mathbf{r_{4,4}}^*$\\
\hline
$2$ & $\mathbf{r_{1,2}}$ & $\mathbf{r_{2,1}}$ &
$-\mathbf{r_{3,4}}^*$ & $-\mathbf{r_{4,3}}^*$\\
\hline
$3$ & $\zeta(\mathbf{r_{1,3}})$ & $\zeta(\mathbf{r_{2,4}})$ &
$\zeta(\mathbf{r_{3,1}}^*)$ & $\zeta(\mathbf{r_{4,2}}^*)$\\
\hline
$4$ & $\zeta(\mathbf{r_{1,4}})$ & $\zeta(\mathbf{r_{2,3}})$ &
$-\zeta(\mathbf{r_{3,2}}^*)$ & $-\zeta(\mathbf{r_{4,1}}^*)$\\
\hline
\end{tabular}
\end{center}
\end{table}

The relay nodes process the received OFDM symbols as given in Table \ref{table_4relay} for which $M=2$, $\mathbf{A_1}=\mathbf{I_4}$, \mbox{$\mathbf{A_2}=\left[\begin{array}{cccc}0 & 1 & 0 & 0\\
1 & 0 & 0 & 0\\
0 & 0 & 0 & 1\\
0 & 0 & 1 & 0\end{array}\right]$}, $\mathbf{A_3}=\left[\begin{array}{ccrr}0 & 0 & -1 & 0\\
0 & 0 & 0 & -1\\
1 & 0 & 0 & 0\\
0 & 1 & 0 & 0\end{array}\right]$ and \mbox{$\mathbf{A_4}=\left[\begin{array}{ccrr}0 & 0 & 0 & -1\\
0 & 0 & -1 & 0\\
0 & 1 & 0 & 0\\
1 & 0 & 0 & 0\end{array}\right]$}. It has been proved in \cite{RaR2} that $\mathbf{C}\mathbf{A_i}=\mathbf{A_i}\mathbf{C}, i=1,2$ and \mbox{$\mathbf{C}\mathbf{A_i}=\mathbf{A_i}\mathbf{C}^*, i=3,4$} for all $\mathbf{C\in\mathscr{C}}$. At the destination node, decoding for $\left\{z_{1I},z_{2I}\right\}$, $\left\{z_{1Q},z_{2Q}\right\}$, $\left\{z_{3I},z_{4I}\right\}$ and $\left\{z_{3Q},z_{4Q}\right\}$ can be done separately in every sub carrier due to the four group decodable structure of $\mathscr{C}$.
\end{eg}

\section{Simulation results}
\label{sec4}
In this section, we study the error performance of the proposed codes using simulations. We take $R=4$, $N=64$ and the length of CP as $16$. The delay $\tau_i$ at each relay is chosen randomly between $0$ to $15$ with uniform distribution. Two cases are considered for simulation: (1) with channel knowledge at the destination and (2) without channel knowledge at the destination. For the case of no channel information, differential encoding at the source as described in Example \ref{eg_4relay_diff} of Section \ref{sec4} is done. When channel knowledge is available at the destination, rotated QPSK is used as the signal set \cite{RaR1}. The transmission rate for both the schemes is $1$ bit per channel use (bpcu) if the rate loss due to CP is neglected.

\begin{figure}[h]
\centering
\includegraphics[width=3.5 in]{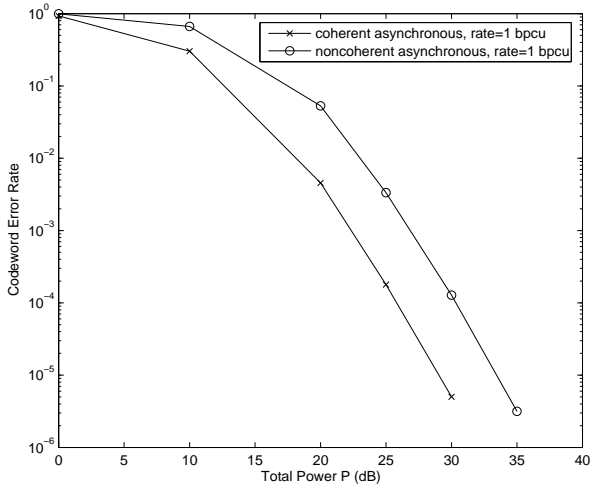}
\caption{Error performance for a $4$ relay system with and without channel knowledge}
\label{fig_simulation}
\end{figure}

 It can be observed from Fig. \ref{fig_simulation} that the error performance for the no channel knowledge case is approximately $5$ dB worse than that with channel knowledge at the destination. This is due to the differential transmission/reception technique in part and also in part because of the change in signal set from rotated QPSK to some other signal set \cite{RaR2} in order to comply with the requirement of scaled unitary codeword matrices. 

\section{Discussion}
\label{sec5}

A drawback of the proposed transmission scheme is that it requires a large coherence interval spanning over multiple OFDM symbol durations. Moreover there is a rate loss due to the use of CP, but this loss can be made negligible by choosing a large enough $N$. In spite of these drawbacks, to the best of our knowledge, this is the first known amplify and forward based transmission scheme available for any number of relay nodes that admits low ML decoding complexity and also provides full asynchronous cooperative diversity. Some of the interesting directions for further work are listed below:

\begin{enumerate}
\item Constructing single symbol decodable distributed space time codes for the proposed transmission scheme.
\item Extending this work to asynchronous relay networks with timing errors and frequency offsets at the relay nodes is an interesting direction for further work. This problem has been addressed in \cite{LiX2} for the case of two relay nodes.
\end{enumerate}

\section*{Acknowledgement}
This work was partly supported by
the DRDO-IISc Program on Advanced Research in Mathematical
Engineering, partly by the Council of Scientific \&
Industrial Research (CSIR), India, through Research Grant (22(0365)/04/EMR-II) to B.S.~Rajan. The authors sincerely thank Prof. X.G. Xia and Prof. H. Jafarkhani for sending us preprints of their recent works \cite{LiX,GuX,JiJ,LiX2}.



\begin{thebibliography}{1}
\bibitem{JiH} Y. Jing and B. Hassibi, ``Distributed space time coding in wireless relay networks," \emph{IEEE Transactions on Wireless Communications}, vol. 5, no. 12, pp. 3524-3536, Dec. 2006.

\bibitem{LiX} Zheng Li and X.-G. Xia, ``A Simple Alamouti Space-Time Transmission Scheme for Asynchronous Cooperative Systems,'' to appear in \emph{IEEE Signal Processing Letters}, Dec. 2007. Private Communication.

\bibitem{GuX} X. Guo and X.-G. Xia, ``A Distributed Space-Time Coding in Asynchronous Wireless Relay Networks,'' to appear in \emph{IEEE Transactions on Wireless Communications}. Private Communication.

\bibitem{RaR1} G. Susinder Rajan and B. Sundar Rajan, ``Multi-group ML Decodable Collocated and Distributed Space Time Block Codes,'' submitted to \emph{IEEE Transactions on Information Theory}. Available in arXiv:0712.2384.

\bibitem{RaR2} ----, ``Algebraic Distributed Differential Space-Time Codes with Low Decoding Complexity,'' to appear in \emph{IEEE Transactions on Wireless Communication}. Available in arXiv:0708.4407.

\bibitem{KiR} Kiran T. and B. Sundar Rajan, ``Partially-coherent distributed space-time codes with differential encoder and decoder,'' \emph{IEEE Journal on Selected Areas in Communications}, vol. 25, No. 2, Feb. 2007, pp. 426-433.

\bibitem{OgH2}  Fr\'{e}d\'{e}rique Oggier, Babak Hassibi, "Cyclic Distributed Space-Time Codes for Wireless Relay Networks with no Channel Information,'' submitted for publication. Available online http://www.systems.caltech.edu/\~{}frederique/submitDSTCnoncoh.pdf

\bibitem{JiJ} Y. Jing and H. Jafarkhani,``Distributed Differential Space-Time Coding for Wireless Relay Networks," to appear in \emph{IEEE Transactions on Communications}. Private Communication.

\bibitem{LiX2} Zheng Li and X.-G. Xia, ``An Alamouti Coded OFDM Transmission for Cooperative Systems Robust to Both Timing Errors and Frequency Offsets,'' to appear in \emph{IEEE Transactions on Wireless Communications}. Private Communication.

\end{thebibliography}
\end{document}